\documentclass[a4paper,final]{appolb}
 \usepackage[]{graphicx} 
\begin{document}
\title{{TRANSVERSE HYDRODYNAMICS WITH SUDDEN ISOTROPIZATION AND FREEZE-OUT}\thanks{Talk presented by WF at the EMMI Workshop and XXVI Max Born Symposium, {\em Three Days of Strong Interactions}, Wroclaw, Poland, July 9 -- 11, 2009, arXiv:0910.0985.}
\thanks{Partly supported by the Polish Ministry of Science and Higher Education, grant N202 034 32/0918.}}
\author{Wojciech Florkowski $^{1,2}$ and Radoslaw Ryblewski $^{2}$
\address{
$^1$ Institute of Physics, Jan Kochanowski University, PL-25406~Kielce,~Poland \\
$^2$ H. Niewodnicza\'nski Institute of Nuclear Physics, Polish Academy of Sciences, PL-31342 Krak\'ow, Poland \vspace{-0.5cm}
}}
\maketitle
\begin{abstract} 
We assume that the early evolution of matter produced in relativistic heavy-ion collisions is described by the transverse hydrodynamics. In this approach only transverse degrees of freedom are thermalized, while the longitudinal motion is described by free streaming. When the energy density of the system  drops to a certain value, the system becomes isotropic (locally, in the momentum space) and freezes out. The sudden isotropization  transition is described with the help of the Landau matching conditions, and the freeze-out is modeled with THERMINATOR. Within this scenario one is able to reproduce in the quite satisfactory way the transverse-momentum spectra, the elliptic flow coefficient $v_2$, and the HBT radii of pions and kaons studied at RHIC (Au+Au collisions at \mbox{$\sqrt{s_{\rm NN}}$ = 200 GeV}). The obtained results indicate that the system produced at RHIC does not have to be fully thermalized in the early stage. On the other hand, the final three-dimensional thermalization is necessary to reproduce the HBT radius $R_{\rm long}$.
\end{abstract}


\medskip

{\bf 1.} Nowadays, the early evolution of matter produced in relativistic heavy-ion collisions is  described most successfully by the perfect-fluid hydrodynamics \mbox{\cite{Kolb:2003dz,Huovinen:2003fa,Shuryak:2004cy}}. Using the appropriate equation of state and incorporating resonances in the modeling of the freeze-out process, the hydrodynamic approach describes very well the hadron transverse-momentum spectra and the elliptic flow coefficient $v_2$. Interestingly, with a suitable modification of the initial conditions, the hydrodynamic approach describes also the HBT radii \cite{Broniowski:2008vp}. 

In spite of its clear successes, the use of the hydrodynamics is still challenged by the problem of early thermalization –-- the correct description of the data requires rather early starting time of hydrodynamics \cite{Bozek:2009ty}, which implicitly assumes a very early equilibration time. A possible solution to this problem was proposed in Ref.~\cite{Bialas:2007gn}. With the assumptions that only transverse degrees of freedom are thermalized and the longitudinal dynamics is represented by free streaming (a physical picture explained naturally in the string models \cite{Bialas:1999zg,Florkowski:2003mm}), one can obtain the {\it parton} transverse-momentum spectra and $v_2$ which are consistent with the data \cite{Bialas:2007gn,Chojnacki:2007fi}. We shall refer to this approach below as to the {\it transverse hydrodynamics}~\footnote{We note that our formulation differs from the formalism of the transverse hydrodynamics introduced originally in \cite{Heinz:2002rs,Heinz:2002xf}. For a more detailed discussion of this issue see \cite{Bialas:2007gn}.}.

In this paper we develop the ideas and formalism introduced in \cite{Bialas:2007gn,Chojnacki:2007fi}. The transverse-hydrodynamics model is supplemented with the sudden isotropization transition followed by freeze-out. This allows for a direct comparison of the model predictions with the {\it hadronic} data. In particular, besides the transverse-momentum spectra and $v_2$ studied already in \cite{Bialas:2007gn,Chojnacki:2007fi}, we analyze now the femtoscopic observables. The isotropization transition is described by the Landau matching conditions, and the freeze-out is modeled with the help of the Monte Carlo code THERMINATOR \cite{Kisiel:2005hn}. We note that THERMINATOR simulates also the hadronization process, since it ``replaces'' fluid elements with the hadronic constituents. 

Certainly, our treatment of the isotropization transition and freeze-out is very much simplified. More elaborate approaches would describe this kind of transformation using kinetic theory or dissipative hydrodynamics, for example, see \cite{Kovchegov:2005az,Bozek:2007di,Zhang:2008kj}. We expect, however, that our scheme may be regarded as an approximation to more advanced approaches where the considered processes have gradual character. 

{\bf 2.} The equations of the transverse hydrodynamics follow from the energy-momentum conservation law, $\partial_\mu T_2^{\mu \nu}=0$, with the energy-momentum tensor defined by the formula \cite{Ryblewski:2008fx}
\begin{equation}
T_2^{\mu \nu} = \frac{n_0}{\tau} \left[
\left(\varepsilon _2 + P_2\right) U^{\mu}U^{\nu} 
- P_2 \,\,\left( g^{\mu\nu} + V^{\mu}V^{\nu} \right)\,\, \right].
\label{tensorT1}
\end{equation}
Here $\tau = \sqrt{t^2 - z^2}$ is the longitudinal proper time and $n_0$ describes the density of transverse clusters in rapidity. The clusters are formed by groups of partons having the same rapidity. They are two-dimensional (2D)  objects, whose thermodynamic properties are described by the 2D thermodynamic variables: $\varepsilon_2$, $P_2$, $s_2$ and $T_2$ (2D energy density, pressure, entropy density, and temperature, respectively). These quantities  satisfy the standard thermodynamic identities: $\varepsilon_2 + P_2 = T_2 s_2$, $d\varepsilon_2 = T_2 ds_2$, and $dP_2 = s_2 dT_2$ (here the baryon chemical potential is set equal to zero). The definition of the energy-momentum tensor (\ref{tensorT1}) includes also the two four-vectors,
\begin{eqnarray}
U^{\mu} &=& ( u_0 \cosh\eta,u_x,u_y, u_0 \sinh\eta), \quad V^{\mu} = (\sinh\eta,0,0,\cosh\eta),
\label{U}
\end{eqnarray}
where $u^\mu = \left(u^0, {\vec u}_\perp, 0 \right)$ is the hydrodynamic flow in the central plane where $z=0$, while $\eta = 1/2 \ln ((t+z)/(t-z))$ is the spacetime rapidity. The four-vectors $U^\mu$ and $V^\mu$ satisfy the following normalization conditions
\begin{eqnarray}
U^\mu U_\mu &=& 1, \quad V^\mu V_\mu = -1, \quad U^\mu V_\mu = 0.
\label{UV}
\end{eqnarray}
The four-vector $U^\mu$ combines the motion of the fluid element in a cluster with the motion of the cluster, thus it corresponds to the flow four-velocity in the standard hydrodynamics.  The term $V^\mu V^\nu$ in (\ref{tensorT1}) is responsible for vanishing of the longitudinal pressure, i.e., in the local rest-frame of the fluid element, where we have $U^\mu = (1,0,0,0)$ and $V^\mu = (0,0,0,1)$, one finds
\begin{equation}
T^{\mu \nu} = \frac{n_0}{\tau} \left(
\begin{array}{cccc}
\varepsilon _2 & 0 & 0 & 0 \\
0 & P_2 & 0 & 0 \\
0 & 0 & P_2 & 0 \\
0 & 0 & 0 & 0
\end{array} \right).
\end{equation}
Very interestingly, such a structure of the energy-momentum tensor appears in the theory of the {\it color glass condensate} and {\it glasma} for \mbox{$\tau \gg 1/Q_s$}, where \mbox{$Q_s \sim 1$ GeV} is the saturation scale expected at RHIC \cite{Kovchegov:2005ss,Krasnitz:2002mn}. This gives further support for consideration of the transverse hydrodynamics as the appropriate  description of the early evolution of matter. 

We solve the equations of the transverse hydrodynamics numerically using the equation of state $P_2 = \varepsilon_2/2 = \nu _g T_2^3/(2\pi)$, where $\nu_g = 16$ \cite{Ryblewski:2008fx}. Our investigations are restricted to the midrapidity region ($z \approx \eta \approx 0$), where the partonic system may be treated as boost-invariant.  The initial conditions assume that the 2D initial energy density at $\tau = \tau_{\rm i} = 1$ fm is proportional to the mixture of the wounded-nucleon density, $\rho_W \left({\vec x}_\perp \right)$, and the binary-collision density, $\rho_B \left({\vec x}_\perp \right)$,
\begin{equation}
\varepsilon_2\left(\tau_{\rm i},{\vec x}_\perp \right) 
= \frac{\nu_g \, T_{2}^{\,3} \left(\tau_{\rm i},{\vec x}_\perp \right) }{\pi} 
\, \propto \,  \frac{1-\kappa}{2}\rho_W \left({\vec x}_\perp \right) + \kappa \rho_B \left({\vec x}_\perp \right).
\label{initcond}
\end{equation}	
The distributions $\rho_W \left({\vec x}_\perp \right)$ and $\rho_B \left({\vec x}_\perp \right)$ are calculated for a given centrality class from the Glauber model. Following the PHOBOS studies of the centrality dependence of the hadron production \cite{Back:2004dy} we take $\kappa=0.14$. The normalization constant required in (\ref{initcond}) determines the 2D initial central temperature of the system, $T_{2 \,\rm i}=T_2(\tau_{\rm i},0)$.

{\bf 3.} The sudden isotropization transition is described by the Landau matching condition
\begin{equation}
T_2^{\mu \nu} U_\nu = T^{\mu \nu}_{3} U_\nu, \label{LMc1}
\end{equation} 
where $T^{\mu \nu}_{3}$ is the standard energy-momentum tensor of the perfect-fluid hydrodynamics
\begin{equation}
T_3^{\mu \nu} = (\varepsilon_3 + P_3) U^\mu U^\nu - P_3 g^{\mu \nu}. \label{LMc2}
\end{equation}  
Here $\varepsilon_3$ and $P_3$ are the three-dimensional (3D) energy density and pressure of the system just after the isotropization transition. Equations (\ref{LMc1}) and (\ref{LMc2}) give
\begin{equation}
\frac{n_0}{\tau} \varepsilon_2 = \varepsilon_3, \label{LMc3}
\end{equation} 
which should be supplemented by the requirement of the entropy growth in the isotropization transition,
\begin{equation}
\frac{n_0}{\tau} s_2 \leq s_3, \label{LMc4}
\end{equation} 	
where $s_3$ is the 3D entropy density. Dividing both sides of Eqs. (\ref{LMc3}) and (\ref{LMc4}) one obtains
\begin{equation}
T_2 \geq \frac{3 \,\varepsilon_3}{2 \,s_3}. \label{LMc5}
\end{equation} 	

\begin{figure}[t]
\begin{center}
\includegraphics[angle=0,width=0.6\textwidth]{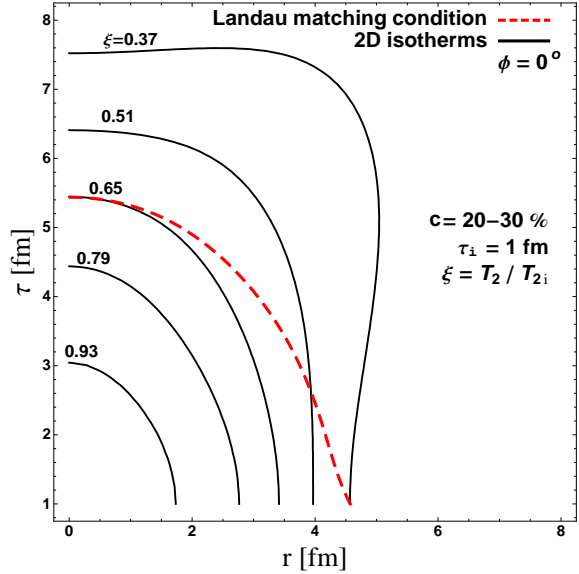}
\end{center}
\caption{\small 2D isotherms describing purely transverse hydrodynamic expansion in the Minkowski space (solid lines). The initial conditions correspond to the non-central (\mbox{$c$ = 20-30\%}) Au+Au collisions at \mbox{$\sqrt{s_{\rm NN}}$ = 200 GeV}. The numbers at the isotherms give the values of the 2D temperature in units of the initial central temperature $T_{2\,\rm i}$. A possible isotropization transition curve defined by the condition $T_{3\,\rm f}$ = constant is represented by the dashed line.}
\label{fig:isotrop}
\end{figure}

In our calculation, $\varepsilon_3$, $P_3$, and $s_3$ are interpreted as the quantities characterizing the hadron resonance gas in equilibrium at the 3D temperature $T_{3 \, \rm f}$. This  means that we identify the isotropization transition with the hadronization process; if the energy density of the system drops to $\varepsilon_3(T_{3\, \rm f})$, the transverse gluonic clusters change into a locally isotropic resonance gas. 

The transverse hydrodynamic evolution of matter produced in the non-central (\mbox{$c$ = 20-30\%}) Au+Au collisions at \mbox{$\sqrt{s_{\rm NN}}$ = 200 GeV} is shown in Fig.~\ref{fig:isotrop}. The solid lines indicate the 2D isotherms defined by the condition \mbox{$T_2$ = const.} The numbers at the isotherms show the values of the ratio $\xi = T_2/T_{2\,\rm i}$. We stress that for massless particles our hydrodynamic equations exhibit scale invariance --- the temperature field may be rescaled by an arbitrary factor without the change of the flow. As described below, this property helps us to fulfill simultaneously the Landau matching conditions for the energy and entropy. The dashed line in Fig.~\ref{fig:isotrop} indicates a possible transition curve defined by the condition \mbox{$T_3$ = const.}  As expected, the line of constant $T_{3}$ does not agree with any of the 2D isotherms. We note that all the curves in Fig.~\ref{fig:isotrop} are plotted for $\phi=0$ (the in-plane direction). Due to the initial asymmetry of the system in the transverse plane, for $\phi \neq 0$ the isotherms have slightly different shapes.

\newpage
Our model has essentially three free parameters: $T_{2 \, \rm i}$, $T_{3 \, \rm f}$, and $n_0$ (the initial time $\tau_{\rm i}$ is always set equal to 1 fm). In the fitting procedure we first assume $n_0=1$ and try to adjust $T_{2 \, \rm i}$ and  $T_{3 \, \rm f}$ in such a way that a good description of the data is achieved. In this step, the energy-conservation condition (\ref{LMc3}) is used as the only constraint. In the next step we rescale the 2D temperature in such a way that the equality $T_2 = 3 \,\varepsilon_3 / (2 \,s_3)$ holds at the initial time $\tau=\tau_{\rm i}$. In this way we make certain that the condition (\ref{LMc5}) is fulfilled on the whole transition hypersurface. The rescaling of the temperature changes the energy density on the 2D side, however, this may be compensated by the appropriate change of $n_0$. In this way we are able to satisfy finally the conditions (\ref{LMc3}) and (\ref{LMc4}).

{\bf 4.} Our description of the isotropization process is quite extreme. Yet, we add to it another extreme assumption: the evolution of the isotropic (locally equilibrated) system is so short that the isotropization process may be followed immediately by freeze-out. Technically, this assumption is realized by the Monte-Carlo simulations done with THERMINATOR, where the 3D transition temperature $T_{3\,\rm f}$ is identified with the freeze-out temperature and the isotropization hypersurface is identified with the freeze-out hypersurface. 

Certainly, our method used to change from the purely transverse hydrodynamic expansion to the isotropic system and further to freeze-out is very much simplified. One may think of several developments, for example, the transition criterion may be changed and/or the time evolution of the isotropic phase may be extended. Such developments are the subject of our current studies. 

\begin{figure}[t]
\begin{center}
\includegraphics[angle=0,width=0.75\textwidth]{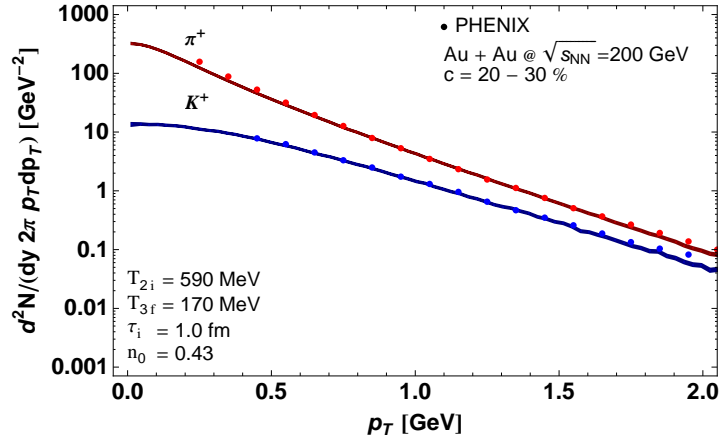}
\end{center}
\caption{\small Model transverse-momentum spectra of pions and kaons (solid lines) compared to the experimental PHENIX data (dots) \cite{Adler:2003cb}. }
\label{fig:figsppt-590-170}
\end{figure}	
\begin{figure}[t]
\begin{center}
\includegraphics[angle=0,width=0.7\textwidth]{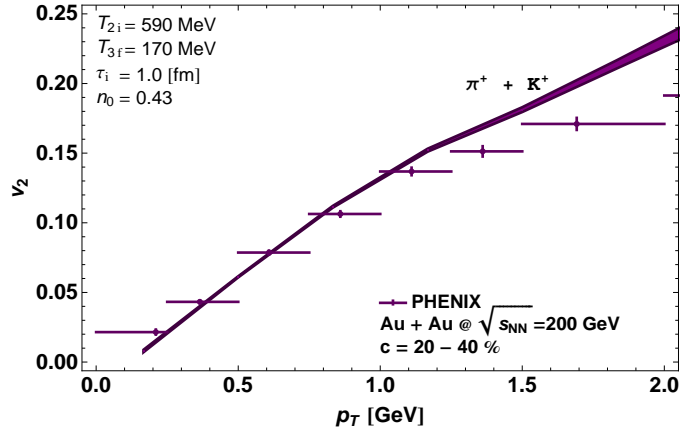}
\end{center}
\caption{\small Elliptic flow coefficient of pions and kaons (solid line) compared to the experimental PHENIX data (dashes) \cite{Adler:2003kt}. }
\label{fig:figv2pt-590-170}
\end{figure}	
\begin{figure}[t]
\begin{center}
\includegraphics[angle=0,width=0.75\textwidth]{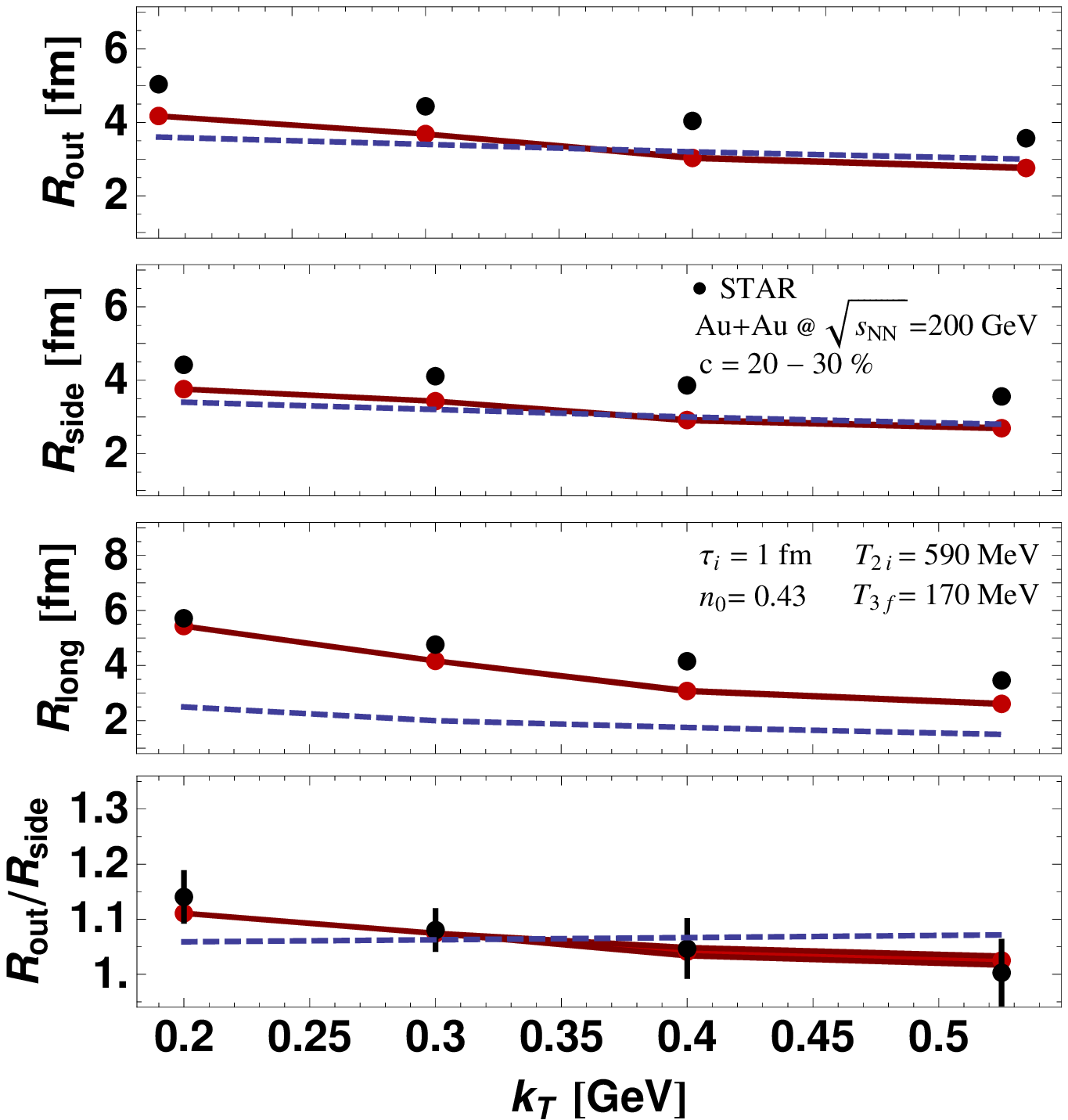}
\end{center}
\caption{\small Model results for the pionic HBT radii (solid lines) compared to the experimental STAR data  \cite{Adams:2004yc} (dots). The radii are plotted as the functions of the average transverse momentum of the pion pair $k_T$. The model parameters are the same as in Figs. \ref{fig:figsppt-590-170} and \ref{fig:figv2pt-590-170}. The dashed lines describe the results obtained without the isotropization transition \cite{wpcf:2008}. }
\label{fig:fighbt-590-170}
\end{figure}	

{\bf 5.} The results of our numerical calculations with the initial distributions corresponding to the non-central (\mbox{$c$ = 20-30\%}) Au+Au collisions at the beam energy \mbox{$\sqrt{s_{\rm NN}}$ = 200 GeV} are shown in Figs.~\ref{fig:figsppt-590-170} -- \ref{fig:fighbt-590-170}. The values of the initial parameters used in the calculations  are: $T_{2 \, \rm i} =$ 590 MeV, $T_{3 \, \rm f} =$ 170 MeV, and $n_0=0.43$. We observe that our model reproduces well the transverse-momentum spectra of pions and kaons, Fig.~\ref{fig:figsppt-590-170}, and also the pion+kaon elliptic flow, Fig.~\ref{fig:figv2pt-590-170}, both measured in Au+Au collisions by PHENIX \cite{Adler:2003cb,Adler:2003kt}. In Fig.~\ref{fig:fighbt-590-170} we show the model pionic HBT radii (solid lines) compared with the experimental STAR data (dots) \cite{Adams:2004yc}. The radii were calculated with the help of THERMINATOR which uses a two particle method with Coulomb corrections \cite{Kisiel:2006is}. The overall agreement with the data is quite good. The theoretical radii are by about 20\% smaller than the experimental values, but their $k_T$ dependence is very well reproduced. Moreover, the model ratio $R_{\rm out}/R_{\rm side}$ matches very well the experimental result. Slightly smaller theoretical values indicate that the real isotropization and freeze-out processes take longer time. This is, of course, an expected feature. 

For comparison, in Fig.~\ref{fig:fighbt-590-170} we also show the HBT radii obtained in the transverse-hydrodynamics model {\it without} the isotropization transition (dashed lines) \cite{wpcf:2008}. We observe that the inclusion of the isotropization transition improves the $k_T$ dependence of $R_{\rm out}$ and $R_{\rm side}$, which leads to the correct behavior of $R_{\rm out}/R_{\rm side}$. The main improvement, however, can be seen in the behavior of $R_{\rm long}$, which for small momenta is increased by a factor of 3.  This means that the full 3D equilibration of the transversally thermalized system is necessary to reproduce the HBT radii. 

{\bf 6.} We conclude with the statement that the transverse-hydrodynamics model supplemented with the sudden isotropization transition and freeze-out is able to describe uniformly the soft pion and kaon data  (transverse-momentum spectra, $v_2$, and the HBT radii).  This brings further evidence that the assumption of the very fast 3D equilibration of matter produced in the relativistic heavy-ion collisions may be relaxed. An earlier study showed that the RHIC data may be explained by the model where the standard hydrodynamic phase is preceded by the parton free streaming \cite{Broniowski:2008qk}. Our analysis of the HBT radii indicates, however, that the ultimate 3D equilibration is necessary to reproduce the experimentally measured value of $R_{\rm long}$

When we turn to heavier particles (protons, hyperons) we reproduce reasonably well their $p_T$ spectra but find difficulties in the reproduction of their elliptic flow --- the model results describing the elliptic flow of protons and hyperons are similar to the pion results, not exhibiting a distinct mass splitting. Thus the model presented here does not explain the HBT puzzle understood as the consistent description of the soft observables for pions, kaons, and (at least) protons. On the other hand, the transverse-hydrodynamics model eliminates the assumption about early thermalization, makes a direct connection to color glass condensate, and describes consistently the pion+kaon sector with very economic tools. A further improvement of the model may be achieved by the matching with the energy-momentum tensor of the dissipative hydrodynamics. 

This investigation was partly supported by the Polish Ministry Of Science and Higher Education grant No. N202 034 32/0918.


\begin{thebibliography}{10}
\expandafter\ifx\csname url\endcsname\relax
  \def\url#1{\texttt{#1}}\fi
\expandafter\ifx\csname urlprefix\endcsname\relax\def\urlprefix{URL }\fi

\bibitem{Kolb:2003dz}
P.~F. Kolb, U.~Heinz, in Quark-Gluon Plasma 3, edited by R.C. Hwa and X.-N.
  Wang (World Scientific, Singapore, 2004), p. 634, nucl-th/0305084.

\bibitem{Huovinen:2003fa}
P.~Huovinen, in Quark-Gluon Plasma 3, edited by R.C. Hwa and X.-N. Wang (World
  Scientific, Singapore, 2004), p. 600, nucl-th/0305064.

\bibitem{Shuryak:2004cy}
E.~V. Shuryak, Nucl. Phys., {\bf A750} (2005) 64.

\bibitem{Broniowski:2008vp}
W.~Broniowski, M.~Chojnacki, W.~Florkowski, A.~Kisiel, Phys. Rev. Lett., {\bf
  101} (2008) 022301.

\bibitem{Bozek:2009ty}
P.~Bozek, I.~Wyskiel, Phys. Rev., {\bf C79} (2009) 044916.

\bibitem{Bialas:2007gn}
A.~Bialas, M.~Chojnacki, W.~Florkowski, Phys. Lett., {\bf B661} (2008)
  325.

\bibitem{Bialas:1999zg}
A.~Bialas, Phys. Lett., {\bf B466} (1999) 301.

\bibitem{Florkowski:2003mm}
W.~Florkowski, Acta Phys. Polon., {\bf B35} (2004) 799.

\bibitem{Chojnacki:2007fi}
M.~Chojnacki, W.~Florkowski, Acta Phys. Polon., {\bf B39} (2008) 721.

\bibitem{Heinz:2002rs}
U.~W. Heinz, S.~M.~H. Wong, Phys. Rev., {\bf C66} (2002) 014907.

\bibitem{Heinz:2002xf}
U.~W. Heinz, S.~M.~H. Wong, Nucl. Phys., {\bf A715} (2003) 649.

\bibitem{Kisiel:2005hn}
A.~Kisiel, T.~Taluc, W.~Broniowski, W.~Florkowski, Comput. Phys. Commun., {\bf
  174} (2006) 669.

\bibitem{Kovchegov:2005az}
Y.~V. Kovchegov, Nucl. Phys., {\bf A774} (2006) 869.

\bibitem{Bozek:2007di}
P.~Bozek, Acta Phys. Polon., {\bf B39} (2008) 1375.

\bibitem{Zhang:2008kj}
B.~Zhang, L.-W. Chen, C.~M. Ko, arXiv:0805.0587.

\bibitem{Ryblewski:2008fx}
R.~Ryblewski, W.~Florkowski, Phys. Rev., {\bf C77} (2008) 064906.

\bibitem{Kovchegov:2005ss}
Y.~V. Kovchegov, Nucl. Phys., {\bf A762} (2005) 298.

\bibitem{Krasnitz:2002mn}
A.~Krasnitz, Y.~Nara, R.~Venugopalan, Nucl. Phys., {\bf A717} (2003) 268.

\bibitem{Back:2004dy}
B.~B. Back, {\it et~al.}, PHOBOS, Phys. Rev., {\bf C70} (2004) 021902.

\bibitem{Adler:2003cb}
S.~S. Adler, {\it et~al.}, PHENIX, Phys. Rev., {\bf C69} (2004) 034909.

\bibitem{Adler:2003kt}
S.~S. Adler, {\it et~al.}, PHENIX, Phys. Rev. Lett., {\bf 91} (2003) 182301.

\bibitem{Adams:2004yc}
J.~Adams, {\it et~al.}, STAR, Phys. Rev., {\bf C71} (2005) 044906.

\bibitem{Kisiel:2006is}
A.~Kisiel, W.~Florkowski, W.~Broniowski, Phys. Rev., {\bf C73} (2006) 064902.

\bibitem{wpcf:2008}
M.~Chojnacki, {talk at IV Workshop on Particle Correlations and Femtoscopy,
  Krak\'ow, Sept. 11-14, 2008, http://www.ifj.edu.pl/conf/wpcf/talks/}.

\bibitem{Broniowski:2008qk}
W.~Broniowski, W.~Florkowski, M.~Chojnacki, A.~Kisiel, Phys. Rev., {\bf C80}
  (2009) 034902.

\end{thebibliography}

\end{document}